# Symmetrical laws of structure of helicoidally-like biopolymers in the framework of algebraic topology. II. *α-helix* and *DNA* structures

## M.I.Samoylovich[1], A.L.Talis[2]

[1]*Central scientific research institute of technology "Technomash", Moscow*
*E-mail: samoylovich@technomash.ru*
[2]*Institute of Organoelement Compounds of Russian Academy of Sciences, Moscow*

## ABSTRACT

*In the framework of algebraic topology the closed sequence of 4-dimensional polyhedra (algebraic polytopes) was defined. This sequence is started by the polytope {240}, discovered by Coxeter, and is determined by the second coordination sphere of 8-dimensional lattice $E_8$. The second polytope of sequence allows to determine a topologically stable rod substructure that appears during multiplication by a non-crystallographic axis 40/11 of the starting union of 4 tetrahedra with common vertex. When positioning the appropriate atoms tin positions of special symmetry of the staring 4 tetrahedra, such helicoid determines an α-helix. The third polytope of sequence allows to determine the helicoidally-like union of rods with 12-fold axis, which can be compare with Z-DNA structures. This model is defined as a local lattice rod packing, contained within a surface of helicoidally similar type, which ensures its topological stability, as well as possibility for it to be transformed into other forms of DNA structures. Formation of such structures corresponds to lifting a configuration degeneracy, and the stability of a state – to existence of a point of bifurcation. Furthermore, in the case of DNA structures, a second "security check" possibly takes place in the form of local lattice (periodic) property using the lattices other than the main ones.*

## 1. INTRODUCTION

In present work it is going to be demonstrated that transferring the above chain of constructions of algebraic topology to a structural level singles out a class of linear biopolymers. The latter means that symmetry laws of structure of linear biopolymers (in particular, α-helices and some forms of DNA-structure) are reduced to realizing constructions of algebraic topology, determining assembly of helicoid-like structures according to a limited number of rules. In other words, the formalism developed in this paper allows one to build *a priori* highly symmetric crystalline structures, which determine symmetry parameters for the mentioned biopolymers.

It is necessary to use 3D lattice constructions corresponding to locally invariant transformations. The problem gets complicated because the problem of classification of 3D manifolds is not only not solved, but it is also not known if there is an algorithmic approach to such a problem [1, v.3, 2-5]. But if one limits himself to consideration of diffeomorphisms (or automorphisms, which in fact is used in consideration of topological structural elements) of the surface, which subtends the 3D structure in question, one may single out certain classes of 3D manifolds. The latter implies a necessity to consider surfaces related to minimal (locally minimal) surfaces and containing singular points which are necessary to bring into correspondence with the selected lattice properties and the automorphisms of the system.

Earlier, upon the derivation of non-integral axes characterizing helicoid-like constructions, [8-15] polytopes have been used as homogeneous spaces. In order to concretize their 3D constructions, let us consider some structural features of a tube-like neighborhood, when in order to embed a manifold (a 2D surface) $M^m \supset R^q$ q>m+1, it is necessary to construct a (q-1) – dimensional smooth manifold which is an N-boundary of tube-like neighborhood. It is built using (q-2) – dimensional disks orthogonal to M, which have as their centers points x∈M, and a sufficiently small radius (neighborhood of the center). The union of such disks is called a tube-like surface of the manifold многообразия M, whose boundary N is mapped in Gaussian manner into $S^{n-1}$. In the case under consideration, $M^2$ is being embedded, so that N must satisfy the Gaussian property for its mapping into $S^3$ which is ensured by projective constructions. Besides, it is essential that the Morse function for N is also a Morse function for M, and the condition for all height functions M to be Morse functions is the regularity of points for the Gaussian mapping $M \to RP^2$. Thus, the use of the considered tube constructions allows one to illustrate the role of disks as manifolds.

There is a general local approach, realized by considering locally minimal surfaces, which allow one to construct associated families of helicoidally similar surfaces. Fixing the ratio of the cylinder diameter and the pitch of the screw in order to ensure zero instability index, allows one to put into correspondence the doubled number of elements on the sphere with the set of elements of double spiral. Minimal surfaces (with zero mean Gaussian curvature) allow one to define conformal mappings on them whereby points on such surface correspond to zeros of derivative function for Weierstrass' representation [3].

The numbering of relations, figures and literature continues and corresponds to the numbers in our previous article "Symmetrical laws of structure of helicoidally-like biopolymers in the framework of algebraic topology. I. Root lattice $E_8$ and the closed sequence of algebraic polytopes".

## 2. SYMMETRICAL LAWS OF $\alpha$-HELIX IN THE FRAMEWORK OF ALGEBRAIC TOPOLOGY.

In the well-known review by A.P. Novikov [19] algebraic topology is defined as "an area of mathematics that appeared when studying those properties of geometric objects that do not change under continuous deformations or maps (homotopies)". Within such an approach it becomes possible to consider simplicial complexes (see Appendix C), corresponding both to a 3D structure of finite volume, as well as its triangulated limiting surface. The closure operation, necessary of obtain a topological space (which is homogeneous) may be simplified by introducing a metric, and a topological group is called discrete if each of its elements allows a neighborhood containing only one point (the identity element of the group is an isolated point).Furthermore, it turns out to be possible to describe a curve, connecting the vertices of the simplexes. The points of such curve, with the exception of vertices of the simplex, do not lie on this minimal surface.

In general, points given by vectors in vector space can be considered by transferring them onto a manifold, and defining them on the tangent (or cotangent) spaces to these points (using diffeomorphisms and differential forms) and thus arrive at using algebraic varieties, given, in particular, by adjoint representations of corresponding transformation groups. Given a Gaussian map $M \to S^2$, putting into correspondence to every point P a unit normal n(P) to a surface M, then such a surface is minimal (locally minimal) given that median curvature valishes: $H \equiv 0$. If M is a Riemannian surface, then the tangent (cotangent) space mentioned above is defined everywhere and is a result of complexification of the real tangent space $T_p$, viewed as a real 2D manifold.

In the complex space (where bounded helicoid and catenoid can be defined), any analytic function of such variables can be considered using the surface given by its zeros. In particular, to define a polytope one uses polynomials with a given system of invariants and vector-valued functions. In the case of the sphere $S^2$ diffeomorphic to $CP^1$ (we are using a transition from $S^3$ to $S^2 \cup S^1$) the projection onto a plane falls into 2 connected parts, and any line on $S^2$, going through 0 and ∞ (in the coordinate cross we get rid of these singular points) also falls into branches that need to be glued in opposite directions.

We use the fact that the 3D Euclidean vector space forms a non-commutative (simple) Lie algebra (with respect to vector multiplication), which is isomorphic to the algebra of quaternions whose norm equals 1, and the corresponding group is locally isomorphic to the rotation group $E^3$. In such an algebra there are no other subalgebras except 1-dimensional ones, and each one-dimensional (taking into account complexity – 2D real) linear subspace forms a subalgebra.

Taking in (12) $\lambda=2$ and correspondingly r=3/4, it is possible to move to a topologically stable helicoid, whose evolvent is shown in fig.6.a. in parallel red lines. Giving the vectors not determining a root lattice, but allowed the algebra of the group of Chevalley type $G_2$, ensures the coincidence of the system of mentioned parallel lines with the system of planes of the group P6mm, an automorphism group of a hexagonal net (fig.7.b). In what follows such an approach allows one to build a helicoid locally periodic structure – a packing of clusters, infinite in one dimension.

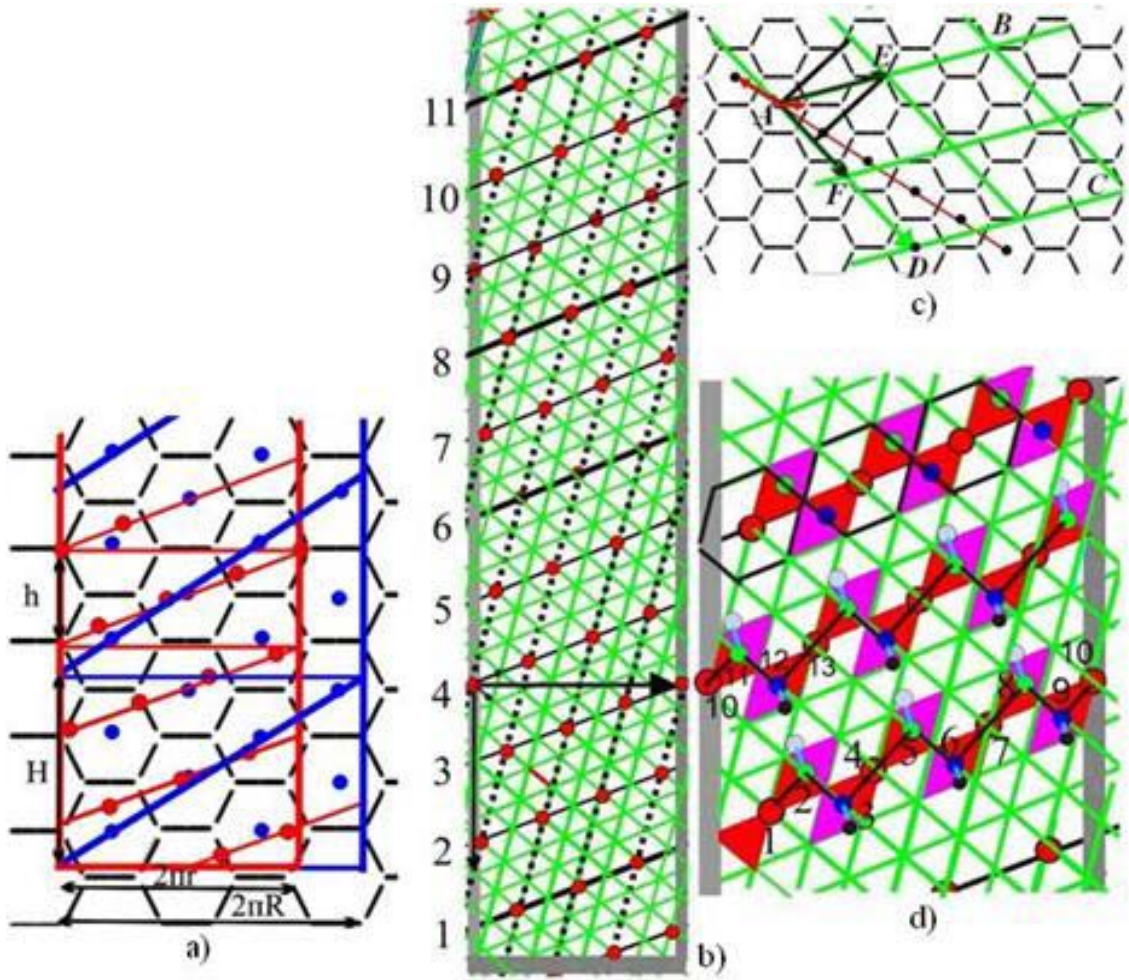

*Fig. 6. a). A hexagonal net, where a strip of the width of 6 unit edges of a hexagon is marked; the blue circles are the centers of regular hexagons. The blue lines, where the centers of hexagons are positioned, constitute the flat development of a helix with the pitch $H \cong 2\sqrt{3}$ (double hexagon height) and the radius R. The red lines within the strip of length $2\pi r=4,65$ represent the flat development of a helix with the pitch h=H/2. There are 40/11 red circles per every turn of the helix.*
*b). The 11 turns of the flat development of the helix shown in a contain 40 red circles, the turns marked by solid lines contains 3 circles each, while the other contain 4. The circles belong to 4 (taking into account the identification of the vertical boundaries of the strip) dashed lines. The flat development may be embedded into the transformed hexagonal lattice shown in green.*
*c). The short $a_1$ and the long $a_2$ vectors of the system $G_2$ are shown as red arrows. The dark green arrows show the vectors with the norms $4a_1$ and $2a_2$ that determine the sublattice (shown in green lines) of the locally transformed hexagonal lattice.*
*d). The red circles are centers of hexagons of the locally transformed hexagonal lattice, between which the pink triangles are situated. They are the common vertices of pairs of red triangles, the midpoints of whose bases contain blue and green circles. The lines joining the blue and green circles of adjacent chains contain white and black circles. The union of the circles, nearest to each other, determines the flat development of a helix, whose 13 consecutive vertices are marked by numbers.*

For the angle of rotation of the helix to equal 40/11, minimal local displacements are necessary of the blue centers of the original hexagons into red vertices on red lines. Then the system of red vertices is cannot be embedded into any of the orbits of 2D subgroups of the group P6mm (fig 6.a,b). A hexagon of the original net is determined by the union of 12 regular triangles of the 2$^{nd}$ coordination sphere of the lattice $A_2$ that determine vectors of the system $G_2$ (fig.4.b). Of the 12 triangles mentioned above, 10 belong to the map $\{3, 6\}^3_{2,1}$, which determines a non-regular 7-vertex partition of a sphere, as well as the union by edges of 4 regular tetrahedra with common vertex (fig 7.a, b, c).

According to [1, 4], this union of tetrahedra represents a special simplicial complex and corresponds to the disk $D^2$ (torus $T^2$). Bringing into coincidence the centers of unions of 10 regular triangles with red vertices on red lines does not lead to partitioning of the strip into regular triangles (fig.7.b.). However, this can be achieved by certain local deformations of triangles (fig.6.d.). Thus, fig.6.d. represents a flat development of a topologically stable, locally periodic packing of simplicial complexes, determined by the polytope {480}, and satisfying all the algebraic and topological requirements listed above.

Upon introduction of a metric (edge length in the chain is assumed to be 1),) for the rod characterized by the axis 40/11 of 4 chains of the 1$^{st}$ type determines a axis with rotation by 99° and the translation h along the axis of the channel. Such a rod substructure may be obtained from the channel 40/10, which is approximated by a helix-like ribbon of regular hexagons (with an edge of length 1), embedded in a cylinder-like surface of radius $r = 3/\pi$. For such helix $h = 2\sqrt{3}$ (fig.6.a), which does not correspond to the value necessary for its topological stability (h/r must be $\cong 2,4$).

Basic parameters characterizing the structure of an α-helix are: the number of amino-acid residues per turn equals 3,6 (in Angstroms), pitch h -5,4, radius r (for cylindrical approximation) – 2,25. Of non-integral axes, giving close values of residues per turn, universally accepted is the axis 18/5, but it does not correspond much to the structure of original cycles. The axis 11/3 (as parameter) is in poor correspondence with the minimal number of residues in domain structure of proteins. Hence the most appropriate is the non-crystallographic axis 40/11. The necessary number of elements per turn is also given by the axis 11/3, but in this case the information about properties of the cycle $3_{10}$ disappears. A helix-like union of 4 cycles $10_3$ gives 40 residues per 12 = 4*3 turns, and the axis 40/11 appears upon decreasing the number of turns by 1. In such formation of the 40/11 axis the $10_3$ cycles are used, representing elements of packings. The total numbers of atoms in such a construction is in correspondence with the polytope {160} (see Appendix E).

Thus, the rod substructure of a polytope characterized by a non-integral axis 40/11, allows one to obtain first a helix of hexagons, embedded in the lattice $A_2$, and then a transition to Chevalley construction for a group of $G_2$-type allows one to obtain from it a local lattice subsystem of atoms $C_\alpha$. Singular points in such a lattice also determine positions of other atoms of the α-helix (fig.6 d). In fact, if one uses standard bond lengths (the double bonds of carbon with nitrogen, carbon and oxygen in these structures are close), a given relationship between radius and pitch, corresponding to the bifurcation point, then it is not possible to construct any other helicoid-like structure. It is necessary to take into account that two variants are possible: either one considers a packing of three atoms C, C' and N given by one surface (with their adjacent atoms), or one uses a system of three tubular (cylindrically similar) surfaces, embedded into each other. The second variant includes the vertices, corresponding to the positions of C, C', N, O and possible positions of atoms H in α-helix (fig. 6.d, 7.e).

In such an approach, the standard symmetry elements are replaced by the reflection operation, when the union of two minimal surfaces possessing symmetry with respect to segments of their boundaries, is also a minimal surface. If the intersection of two smooth minimal surfaces $M_1 \cap M_2$ contains some open subset, then their union is also a smooth minimal surface. If $M_1$ and $M_2$ are minimal ruled surfaces and possess congruent frames (a family of generatrices everywhere dense on the surface), they also are congruent.

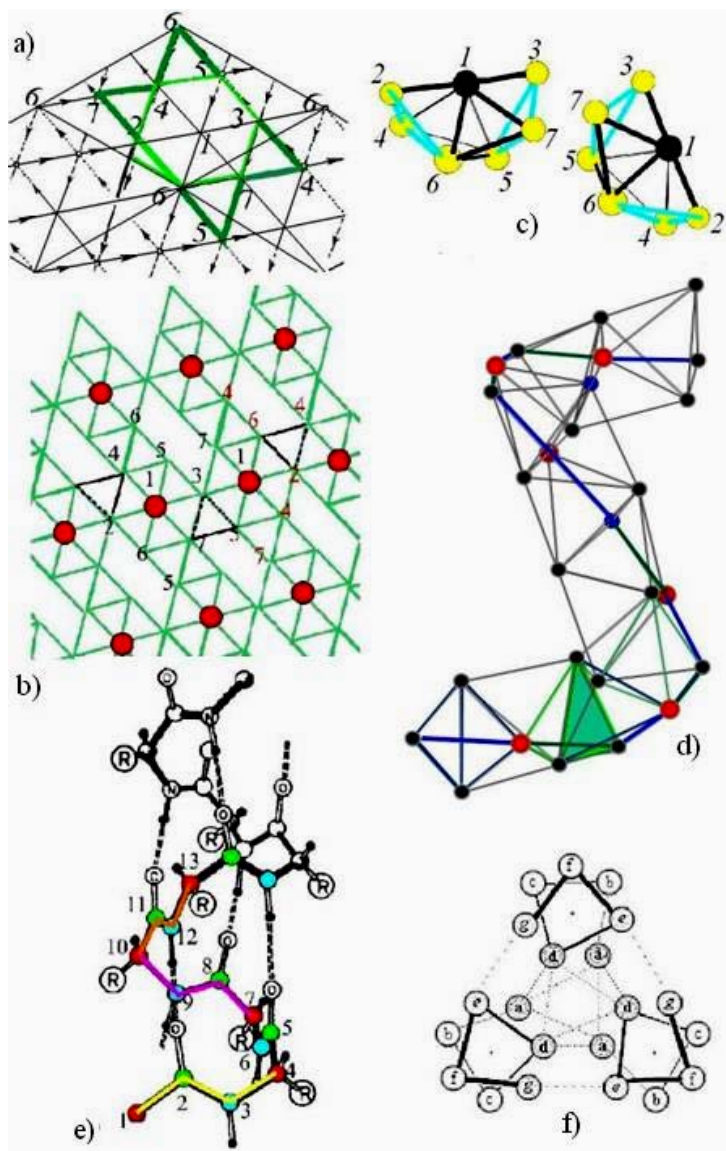

*Fig.7. a), c) The map $\{3,6\}_{2,1}$, where 10 triangles of the non-regular map $\{3,6\}^3_{2,1}$ are marked. Identifying the vertices with equal numbers determines a 7-vertex union (face -to-face) of 4 regular tetrahedra with the common vertex 1. Two such 7-gons may be united (as manifolds) by their common face 3-5-7.*

*b) Every red circle of the flat development(fig.6.d). is the vertex 1 in the union of 10 triangles of the corresponding map $\{3,6\}^3_{2,1}$. The numbers of vertices of triangles coincide with numbers on fig.7.a. The triangles 3-5-7 are the pink triangles on fig.6.d.*

*d) The substructure of 7-vertex unions of quadruples of tetrahedra (collected by their faces of the type 3-5-7), which is determined by identifying equal vertices of the flat development fig.7. b. Every 7-vertex union contains two "exterior" tetrahedra, shown by their blue and green edges. The blue and green tetrahedra in the substructure have a common blue-green edge. The blue and green edges shown in solid lines join the red centers of the quadruples of tetrahedra and form a spiral, characterized by the axis 40/11. The blue and green circles of the flat development on fig.6.d. correspond to blue and green spheres in blue and green tetrahedra.*

*e) The α-helix[21,22] as a realization of the flat development of fig.6.d as well as the union of tetrahedra fig.7.d: red, blue, green, white and black balls depict the atoms $C_\alpha$, N, C', O and H. The colors and numbers of atoms are the same as in fig 6.d.*

*f) The super-helix of 3 α-helices [22] as a mapping of the structure of polytope {480} (fig.3.e.): inside the triple of rods, characterized by the axis 40/11(fig.5.c) and corresponding to the α - helices, there appears a channel 30/11(fig.5.a), where the rods 40/9 lie between the pairs of rods 40/11 (fig.5.b).*

Structural features of biological (primarily protein) structures contain certain information concerning the mathematical formalism that ought to be used in their description. For instance, the presence of bases for coding the proteins, three-atom chains like $C'$-$C^\alpha$-N, the characteristic $3_{10}$ cycle and others. For an α-helix, in fact, one considers an ordered locally-periodic rod packing of cycles, where everything said above is applicable both to a chosen point of the molecular cycle ($C^\alpha$ in the given case), and to the other atoms. Locality (finite nature) corresponds to local subsystems of the root lattice $E_8$ (as well as $A_2$) and to the helicoid character (axes 11/3 and 40/11), which ensures the minimality of integral energy as well as stability of the system with respect to external forces (small but finite).

## 3. TOPOLOGICAL AND ALGEBRAIC STRUCTURAL FEATURES OF FORM DNA STRUCTURES.

The structure of DNA, formed by unordered according to laws of crystallography large molecules, is impossible to describe in terms of atomically generated lattice, and therefore, it is necessary to use the variant of local-lattice packing. In DNA structures takes place a transition from local lattice (locally periodic) atomically generated structures to locally periodic packings of molecules. Scale invariance of the system (a most general form of fractal transformations) is put into action by a peculiar local conversion, when repeatability in chains and the number of elements transform into a characteristic of the axis of a helicoidally-similar rod and then into elements of helicoidally similar local lattice packing.

A more complicated situation exists with molecule packings in DNA structures of various forms форм [21-23], where elements of repeatability for various combinations in such packings manifest themselves in well-known coding specifics. The differences in atomically-generated local lattice structures of such type, as well as in analogous local lattice packings consists in that in the first case the numerator of the non-crystallographic axis characterizing cylindrically similar substructure gives unitary periodicity for a given kind of atoms via the number of such atoms. While in the packings the parameters of such a non-crystallographic axis carry, in known degree, a formal sense, showing the number of positions with analogous properties to situate packing elements, represented by molecules. This does not imply, however, that such properties of atoms, making up the packing, as being cyclic, will not appear in structural parameters.

The sphere $S^3$ may be given not only as a union of the form $(S^1 \times D^2) \cup (D^2 \times S^1)$ with Rib's foliations, using tori $T^2$ and the boundary torus, but also as a manifold SU(2). In such case, using unitary periodicity and the fact that $S^7$ also gives the principal fiber space for SU(2), one may also give in a necessary form (and according to invariants) the disk $D_0^2$. It can represent a cut of $S^7$ by a 3D plane which is a geodesic manifold in the group SU(2), remaining such also under rotations of its boundary circle $S^0$. In a discrete version this implies a rotation by certain angles corresponding to the given group as well as the mapping $D^2 \to SU(2)$ with fixed $S_0 \to SU(2)$ in the case of isometric transformations (motions). Such an approach allows one to put into correspondence the set of midpoints of minimal geodesics, containing (in the space of paths) the points corresponding to identity transformations, with the Grassmanian manifold $G_{2,1}$ (the set of unitary complex matrices), and with the intersection of the group and its algebra SU(2)∩su(2), realized in $E^4$ by complex matrices 2×2.

In a general form (see Appendix A), for minimal surfaces their stability is determined with respect to changes in their surface area (as well as the volume enclosed by it) under small deformations. Stability is characterized by [3] the index Ind M, which corresponds to the number of ways to change the surface area. If this index is not zero, the surface M is unstable. The indices of all periodic minimal surfaces (including the infinite helicoid) are infinite and are, in the said sense unstable [3]. However, there are well developed methods to construct complete minimal surfaces, embedded in $E^3$, using Weierstrass' representations, obtained using a 1-form, holomorphic on M and determined by the so called well meromorphic functions. The latter include all meromorphic functions with finite number of zeros (or poles), all fractional rational, trigonometric and hyperbolic functions.

In reality, local representation of minimal surface is reduced to using solutions of the form above, when $\varphi=\partial r/z$ corresponds to the holomorphic radius-vector $(\varphi)^2=0$ (the condition for components of a holomorphic function, for surface regularity in conformal coordinates, corresponding to defining nilpotent groups and, in particular, p-groups and corresponding p-algebras). The Gaussian map itself (tangent map) preserves angles between vectors under diffeomorphisms. It is shown [3] that Weierstrass representations allow one to define catenoid as well as complete helicoid, and, in general case, an associated family for some minimal surface M (for instance, of helicoid or catenoid) consists of locally isometric minimal surfaces (incongruent pairwise, as a rule).

Let S be some surface given by Weierstrass' representation $(U,\omega, (aw+b)^m))$ [3], $a,b \in \mathbf{C} \neq 0$, where $w=u+iv$ gives via isothermal coordinates $(u,v)$ the parameters of a representation after a period (an integral over a closed piecewise-smooth curve, not retractable by continuous deformations to a point). For the case in question of unitary periodicity m is a non-zero integer, $\omega$ is 1-form, $U \subset \mathbf{C}$ is a subdomain in the complex plane. The surface S is characterized by zero index in two cases, namely, in the case of the image of a domain U under a Gaussian mapping, and for m=1 in some open subset $K^*=S^2 \cap (x^3 \leq 0)$. This subset may be defined for an open hemisphere $S^2$ without pole or it may be a part of the sphere $S^2$, enclosed between two parallel planes, separated from the center of the sphere by the distance th $t_0$, where $t_0$ is the only root of the equation that determines the only positive root of the equation (12): cth $t_0 = t_0$ [3], corresponding to the bifurcation point. Note that introducing "unused" vertices on $S^2$ (in the neighborhoods of poles, the number of which, under the above conditions is 1/6 of the original) corresponds to a transition from a 96-vertex polyhedron to a 80-vertex one, and for the polytopes {1152} and {576}, respectively, to the 960 and 480–vertex manifolds.

As is known, all non-planar non-congruent complete minimal surfaces form a one-parameter family of helicoids or (for surfaces of revolution) one-parameter family of catenoids, which ensures, with appropriate values of their parameters, a local diffeomorphism (preserving the metrics) of both configurations. The parameters of such families can be chosen as the pitch (h) for helicoids, and the radius (r) for catenoids. The difference between these configurations is that while the generatrices of catenoids are catenaries ν(u) in the plane XZ (so that the coordinate Z of points on ν(u) grows with increasing u), for the helicoid it is the Y axis (so that with increasing u the Y – coordinate decreases).

The median surface curvature for a helicoid is zero along any generatrix, so that for any pair of generatrices there is a motion, transforming one generatrix into another, and the helicoid into itself. Under certain conditions one can constructs a configuration, unifying helicoid and catenoid, namely, to build a system of summary radius-vectors for both configurations and equal values of u. Here (using complex variable $w=u+v$ in $\mathbf{C}$) the radius-vectors $\mathbf{r}(u,v)$ are used, where v corresponds to the angle $\varphi$ in the cylindrical system of coordinates. Here it is necessary to draw a catenary ν(u), where it is possible to make ν(u) discrete by partitioning into a system of piecewise smooth segments, a family of lines parallel to the Y axis, so that the points u=0 correspond to the vertices of such chain (in this case it is possible to use edges and vertices of polytopes).

The surface thus obtained is characterized by the shift |u| as well as by some curve ω, whose points move to the right with increasing u, and to the left with its decrease. At the same time we notice that when the height of the generatrix of the helicoid changes by the pitch h, the curve ω is lifted by the same amount. This family of lines is supplied on the surface by giving the radius-vector of the form [3]:

$$\mathbf{r}(u,v, \alpha)=\cos\alpha\, \mathbf{r_1}(u,v) + \sin\alpha\, \mathbf{r_2}(u,v) \ ,$$

where $\alpha \in [0,\pi/2]$; for $\alpha=\pi/2$ the curve ω turns into a generatrix of the helicoid, and for $\alpha=0$ into the catenary of the catenoid.

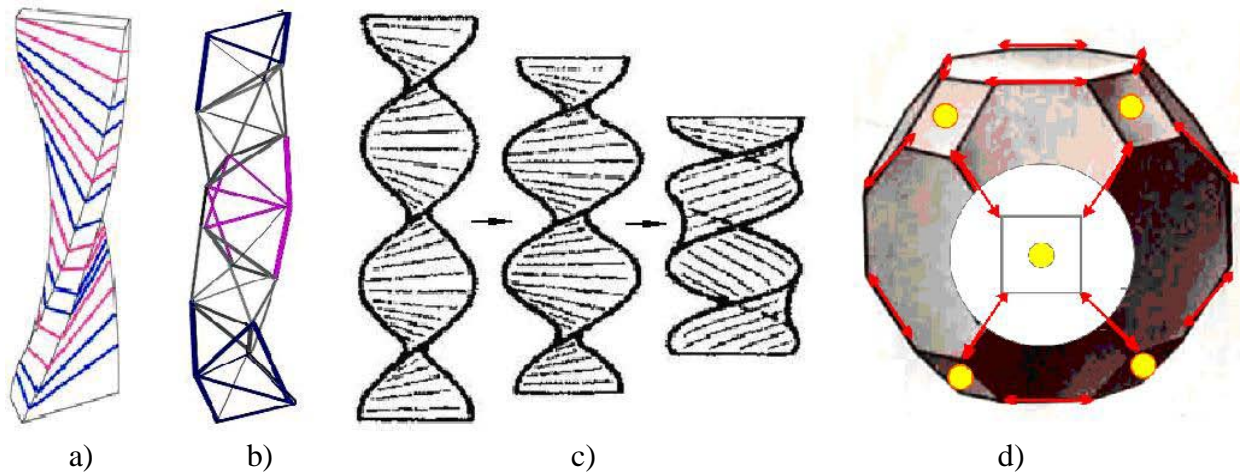

a)    b)    c)    d)

*Fig. 8. a) Union of rods by law helix.*
*b) Partitioning the rods (shown in blue and pink lines) into tetrahedra, determines their uniting (as manifolds) by a connected sum [1], which is represented by three gray tetrahedra.*
*c) A closed path consisting of two helices and two closing segments. Given a large pitch of helices a soap film, put over such a path, is a physical realization of a helicoidally similar structure; diminishing the pitch of the helix leads to a transformation of a helicoid into a double spiral helicoid-like surface of the catenoid, uniting the mentioned subsystems [3].*
*d) Discarding the polar caps of the sphere leads to formation of an equatorial belt on it. In a discrete version of a 48-vertex polyhedron (fig.3.e.), a 40-vertex manifold is selected in this way.*

This approach allows one [3] not only to construct a surface, combining properties of the helicoid and the catenoid, but also form a configuration out of their fragments, determined by two helices and two closing segments. Changing parameters of such configurations, they can be transformed by isometric transformations into conjugates, turning into minimal surfaces of intermediate type or into associated surfaces. It is possible that these processes are related to separation of the two-spiral configuration into two one-spiral ones followed by recovery of every spiral into a stable double spipral. Changing parameters of such configurations, one may bring the surface to a minimal surface of intermediate type by isometric transforms. It is possible that such processes are related to the separation of the two-helix configuration into two single-helix ones with following recovery of each helix into a topologically stable manifold. In the case of a sphere $S^2$ diffeomorphic to $\mathbf{CP}^1$ (using the transition (4) from $S^3$ to $S^2 \cup S^1$), its projection on the plane falls into two connected parts, and any line on $S^2$ going through 0 and ∞ (we discard the abovementioned singular points in the coordinate cross) also falls into branches that must be identified upon gluing in opposite directions.

Existence of a sub-period (divisibility of a turn into thirds) is in correspondence with existence of three non-unit involutions in the Chevalley group of type $G_2$. If in an α-helix the construction process uses the handle as a topological operation, then in the structure of DNA it is not possible to describe construction without operations of connected summation (more on topological structural elements see in [1, 15]). The construction of the double helix DNA structure may (with the many restrictions mentioned above) be viewed as some analog of the α-helix with the triple period. In fact, if for DNA we have the period of 35 A with the diameter about 19 A, then formal parameters of one spiral (outside the construction) may be viewed as represented by values of 8-9 A for the diameter 2r, and 11-11,5 A for the pitch h, respectively. In such case the ratio h/r corresponds to the above value for the bifurcation point for a separate helix. Thus, if an α-helix as a rod substructure is characterized by a 40/11 axis, then the DNA structure in this approximation may be characterized by a formal axis (as parameter) 120/11.

Action of groups of homologies (cogomologies) does not depend upon the triangulation of the manifold and is therefore homotopically invariant. At the same time, homotopy of maps is related to coincidence of homomorphisms of their tangent subspaces, and, therefore (under certain conditions) also for corresponding algebraic constructions. Simplicial homologies (as a particular case of cell homologies) are also homotopically invariant. The Euler characteristic ($\chi$) is a homological as well as a topological invariant of the surface. Therefore, $\chi$ is a kind of a scale-invariant characteristic of the topological volume. This invariant of a manifold (peculiar bridge to subsequent fractal constructions) is a scale-invariant characteristic of volume, enclosed by the manifold, however, using the said characteristic is related to various difficulties [1-2].

In fact, when defining local lattice structures, one should take into account that the lattice (in algebraic meaning) is not necessarily defined as a subgroup of the n-dimensional real space, generated by n linearly independent vectors (as in the case of root lattices). One may also use (see Appendix F) complex and quaternion vectors, because in addition to the integers Z, there are three rings of whole numbers, namely: Gaussian (complex integers), Eisenstein and Hurwitz numbers (quaternion integers), used by us before. The root lattice $D_4$ (or root system $F_4$) is used in defining polytopes on $S^3$, for instance, in Gosset's construction for {3, 4, 3}. Along with it, one may also use a 2D Gaussian lattice (the A construction [4]), whose minimal vectors have the form ($\pm 1, \pm 1$), ($\pm 1, \pm i$), ($\pm i, \pm i$), ($0, \pm 1 \pm i$), (as well as lattices, related to treating the Hurwitz group as a multiplex group $2A_4$), and realification again leads to the lattice $D_4$ [4].

In the theory of phase transitions of the 2$^{nd}$ type (PT-2 [20]) the group of the wave vector **k** contains rotational axes, for which the conditions of invariance with respect to translations for basis functions and representation sets must hold. Using vector representations in the description of local phase transitions automatically lifts this assumption, replacing it with others. For instance, using manifolds, put into correspondence with root lattices, assumes equivalence of two types of vectors ($\pm \alpha$), which corresponds to introduction of central symmetry in the usual treatment of PT-2 (or to the doubling of the number of vectors in a star in the absence of central symmetry). However, for the constructions in question, the analogy is in the requirement of integrality of vector coefficients, which is satisfied in the method of reduced Brillion zone (RBZ), with an n-fold increase in lattice cell parameters. In the self-dual lattices, one may also use the variant of reduced elementary cell (REC) for the 1$^{st}$ and 2$^{nd}$ coordination spheres, while increasing n-fold the corresponding parameters of the reciprocal lattice.

In the variant in question it is possible to represent [5, 17-18] the projection of the polytope {3,4,3}, whose vertices are the vectors from 24-element classes. A fig. 9.a shows a projection of the polytope {3,4,3}, whose vertices are put in correspondence with elements of such non-principal lattice (see Appendix F). The polytope {3,4,3} is a cell of a honeycomb {3,4,3,3}. Hence such a honeycomb can be projected onto a plane into the union of its projections, which is a partition into 5, 6- and 7-vertex figures. This partition corresponds to 81 class of the factor-manifold $D_4/3D_4$ of the self-dual lattice $D_4$, which (except for zero vector) is represented by 24 classes of unit vectors of the norm [2], 24 classes of unit vectors of the norm [4], and 32 classes, each of which consists of three vectors of norm [6]. In the case in question one may set apart the vectors related to points obtained in this way and consider their relationships with conjugation classes of the lattice for $D_4/nD_4$.

An additional requirement (for the conditions listed above) is that three edges meet in one vertex (on a plane), as shown in fig.9 c), d). We obtain a system of points characterized by a 12-element subgroup that can also be viewed as a cut. This system can be put into correspondence with the system of elements of a dodecagonal quasicrystal as well as with the rods defined above.

The discussion above allows one to consider a 96-vertex polyhedron and a 80-vertex manifold selected in it (fig.3.f, fig.10.a) as a map (in the form of the mentioned union) of the polytope 1152, in which a 960-vertex manifold has been chosen. Then every vertex of the polyhedron corresponds to 12 points on $S^1$. In the partition of the sphere, determined by the polyhedron, into 5, 6, and 7-gons, an equatorial strip is uniquely selectable of alternating 5- and 7-gons, which can be developed into a helicoidally-similar strip (fig. 10.a.). Viewing every 5- and 7-gon as a cross section of a rod obtained as a union of lateral hexacycles, then the tape can be put into correspondence with a helicoidally similar union of rods, whose ends form a double spiral. The union of such rods obtained according to fig. 10.a is presented in the form of a diamond-like structure on fig. 11a.

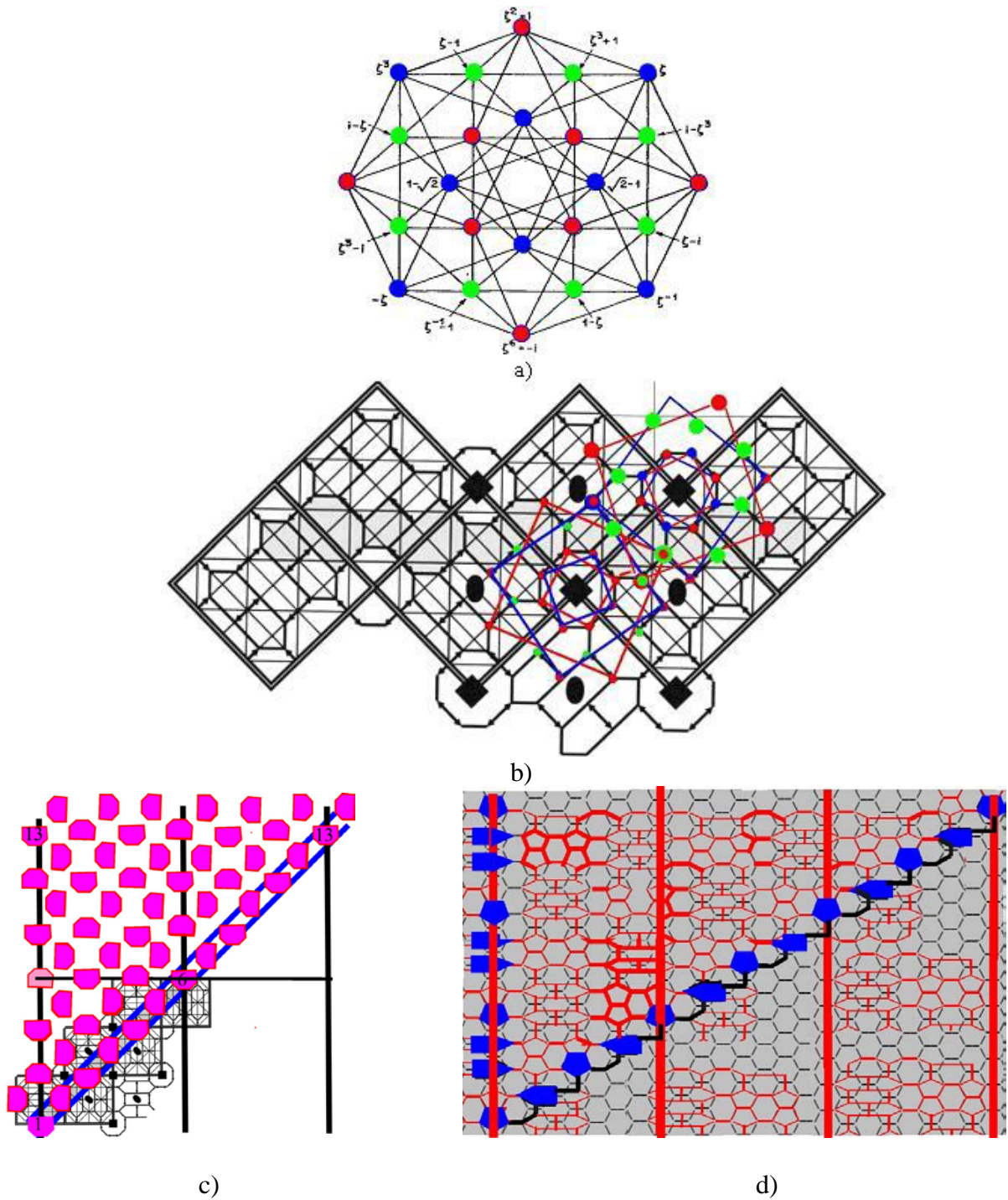

Fig.9.*a) A projection of the polytope {3,4,3}, whose vertices are represented as elements of a non-principal lattice [5]. The coloring shows a partitioning of 24 vertices into 3 orbits of 8-element cyclic group and into 6 orbits of its 4-element cyclic subgroup.*
*b) Equal-edge 3-coordinated partitioning of the flat development of a cube whose vertices are determined as unions of projections a). Two adjacent projections intersect over 4 vertices; every face of the cube contains 16 vertices of the partition. The midpoints of the edges, forming the Petri polygon of the cube, contain 7-gons.*
*c) In the partitioning b) of the flat development of a cube a strip of alternating 5- and 7-gons has been selected. Vertical lines divide the strip into parts, containing 6 edges of the Petri polygon of the cube. The 13$^{th}$ 7-gon of the strip is on the 3$^{rd}$ vertical line and may be identified with the 7$^{th}$ 7-gon, situated on the 1$^{st}$ vertical line when turning a flat development into a cylinder-like surface.*
*d) Discarding the common edge of two squares of a square net of the partition c) leads to an equal-edge, 3-coordinated partition of the hexagonal net. In the strip of alternating 5- and 7-gons shown is only the zigzag-like union of 5-gons. The 1$^{st}$ and the last vertical lines on c) and d) coincide.*

According to the previously considered sequence of constructions of algebraic topology this union possesses a 12-fold non-crystalloid axis and can be put into correspondence with the structure of Z-DNA. (fig.11.b, c). Within the approach being developed a transition from the Z-form of DNA to the B-form corresponds to a transition from the mentioned polyhedron {96} to a 96-vertex partition of the sphere possessing the 5th axis.

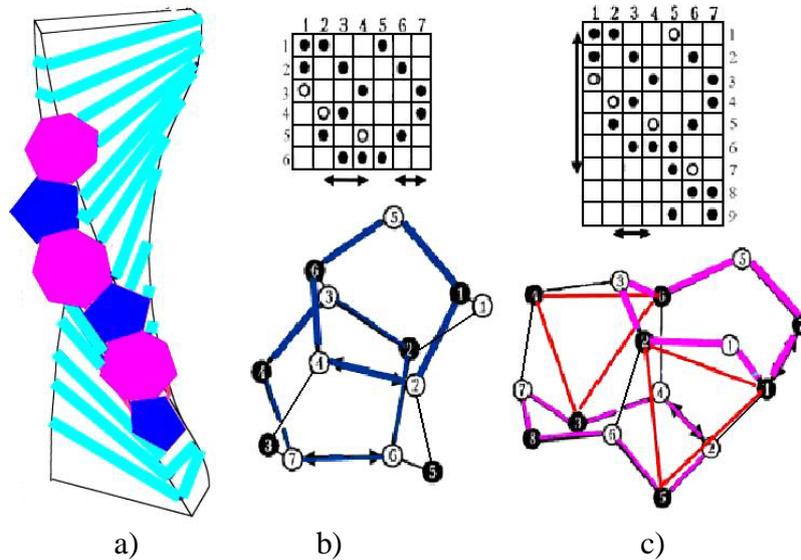

a)                b)                c)

*Fig.10 a) Helicoid –like union of rods with transversal blue pentacycles and pink heptacycles, which is determined by the flat development of the equatorial belt (alternating 5- and 7-gons) of the polyhedron {96} (fig.3. f).*
*b), c). Special clusters of diamond-like structures determined by the flat developmt of fig.3.f. as well as the incidence tables of finite projective geometries [6-7]. In the cluster c) the upper and lower red triangles correspond triangles from connected sum (see fig.8. a. b.).*

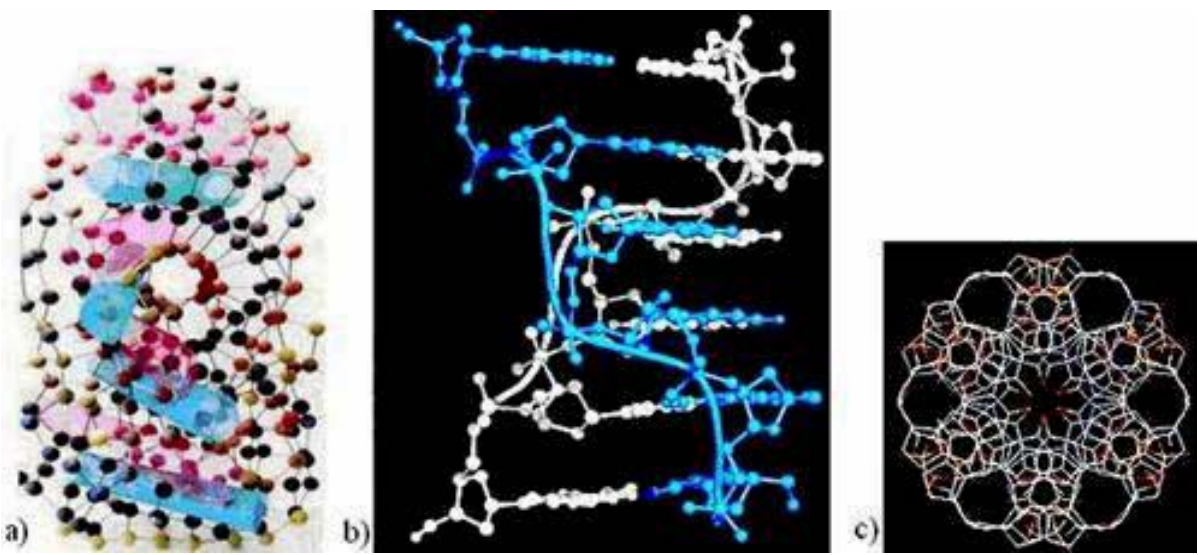

*Fig.11. a) A model of diamond-like structure [11], containing a helicoid-like union of rods with transverse penta- and heptacycles with inserted blue and pink tubes. All the atoms are 4-coordinated, rods are formed by side hexacycles. The pentacycles limiting blue rods form two helicoid-like systems.*
*b) A model of Z-DNA as a union of linear structures of bases by a non-crystallographic axis of 12[th] order [22,23]. Shown are 6 linear structures of base pairs, whose ends form two helices of blue and gray pentacycles joined in a zigzag-like fashion. Each linear structure of this kind corresponds to a blue tube in a).*
*c) A rear view of the Z-DNA [23].*

## 4. CONCLUSION

The treatment given above shows that the α- helix as well as Z-DNA structures are described within the framework of algebraic topology as special local lattice packings on helicoid-like surfaces. Volumes of such systems are enclosed by surfaces corresponding to bifurcation points for minimal surfaces of helicoid-like type, given by Weierstrass' representation and satisfying the requirement that the index of instability of the surface equals zero. In other words, the developed approach shows that the necessary condition of stability and reproducibility of the biological structures in question is their correspondence to a unique system of constructions of algebraic topology. It determines assembly of atoms (molecules) according to topological properties of the real physical world and the condition, that finite discrete ordered structures can be embedded into it. As is predicted by the catastrophe theory [16], formation of such structures corresponds to lifting a configuration degeneracy, and the stability of a state – to existence of a point of bifurcation. Furthermore, in the case of DNA structures, a second "security check" possibly takes place in the form of local lattice property using the lattices other than the main ones.

Modern X-ray structural methods do not always give adequate information concerning local periodicity of live cellular structures. In fact, their rod substructures, corresponding to crystallographic axes outside crystalline structure, may be logically characterized by non-crystallographic, mainly non-integral axes. Correspondingly, ignoring local periodicity leads to formal breakdowns of the basic paradigm of biological reductionism [24], because cases have been experimentally observed when different proteins of the same protein with varying functions were coded by the same gene with certain sequence of bases. This irregularity could have been ignored, were it not for a characteristic partition of the original DNA helix into separate subsystems, for instance, when reading the information. In this process the active and passive elements may change the effective length of the corresponding RNA molecule, which leads unavoidably to a change in type of the non-integral axis that characterizes such structure, as well as to certain structural features of macromolecules for the entire chain. A consequential consideration shows that it is the B-form [25] that is closest to the DNA structure in the living cell.

DNA does not only contains the necessary functional code, but also realizes a very important transition for the local-lattice (locally periodic) atomically generated structures to local-lattice (locally periodic) packings of molecules and further size increases of coded systems. Scale invariance of the system (the most general type of fractal transformations) is set into action by a specific local conversion (transition), when the repeatability (local atom-based lattice) in chains and in the number of elements turn into a characteristic of an axis of the helicoidally similar rod and then into the elements of the helicoidally similar local-lattice packing.

## References

1. B.L. Dubrovin, S.P. Novikov, A.T. Fomenko. Modern Geometry Éditorial URSS, Moscow, 2001. [in Russian].
2. A.G. Kurosh. Theory Groups. Moscow. Nauka.1953. [in Russian].
3. A.T. Fomenko, A.A. Tuzhilin. Elements of the geometry and topology of minimal surfaces in three-dimensional space. Translations of Mathematical Monographs, 1992. V.93.
4. J. Humphreys. Linear Algebraic Groups. New York, Berlin: Springer-Verlag. 1975.
5. J.H. Conway, N.J.A Sloane. Sphere Packing, Lattice and Groups. New York, Berlin, Heidelberg, London, Paris, Tokyo: Springer-Verlag 1988.
6. H.S.M Coxeter. Regular polytopes. N.Y. Dauer, 1973.
7. H.S.M Coxeter. // Philos. Trans. R. Soc. London , Ser. A 1930, V.229, P. 346.
8. M.I. Samoylovich, A.L. Talis. Algebraic Polytopes and Symmetry Laws of Ordered Structures. //Doklady Physics, 2008, V. 53, No. 6, P. 292–297.
9. M.I. Samoylovich, A.L .Talis. Special Class of Helicoids with Crystallographic, Quasicrystallographic, and Noninteger Axes //Doklady Physics, 2007, V. 52, No. 5, P. 247–252

10. M.I. Samoylovich, A.L. Talis. Transformations of Gosset Rods as the Structural Basisof Gas-Hydrate–Ice Phase Transitions. //Doklady Physics, 2009, Vol. 54, No. 4, P. 161–166.

11 M.I. Samoylovich, A.L Talis. A foundation for the theory of symmetry of ordered nanostructures. Moscow: CNITI "Technomash", 2007.

12. M.I Samoylovich, A.L Talis. Symmetrical features and local phase transitions of ordered solid structures: tetravalent structures of gas hydrates. //Crystallography Reports 2009, V.54, №7. P.1101-1116.

13. M.I Samoylovich., A.L Talis. Gosset helicoids: II. Second coordination sphere of eight - dimensional lattice $E_8$ and ordered noncrystalline tetravalent structures //Crystallography Reports 2009. V.54, №7 P.1117- 1127.

14. M.I Samoylovich, A.L Talis. A special class of simple 24-vertex polyhedra and tetrahedrally coordinated structures of gas hydrates. //Acta Cryst. A, 2010, V.66, P. 616-625.

15. M.I. Samoylovich, A.L. Talis. Symmetrical laws of structure of helicoidally-like biopolymers (a-helix and DNA-structure) in the framework of algebraic topology. //Proceedings the XVIII International Conference "High Technologies in Russian Industry" Moscow: CNITI "Technomash P. 394-424. 2012. [in Russian].

16. V. I .Arnold. Catastrophe theory. Springer-Verlag, New York, 1984.

17. M.Koca, R.Koca, M.Al-Barwani, J. Math. Phys. 2006. 47, 043507-1.

18. F J Sadoc, N. Rivier Boerdijk-Coxeter helix and biological helices //Eur. Phys. J. 1999. B12, P.309-318

19. S.P.Novikov. Algebraic topology. //Modern problems of mathematic/. 2004, V. 4, Moscow:. 46 p. [in Russian].

20. L.D.Landau and E.M.Lifshitz. Statistical Physics 1980. Vol. 5 (3rd ed.).Butterworth-Heinemann.

21. G. E. Shulz, R.H. Schirmer. Principles of Protein Structure. Springer-Verlag New York-Heidelberg-Berlin, 1979

22. Finkelstein A.V., Ptitsyn O.B. Protein Physics. Academic Press, An Imprint of Elsevier Science; Amsterdam - Boston - London - New York 2002.

23. Lehninger Principles of Biochemistry (5th edition) By David L. Nelson and Michael M. Cox Publisher: W. H. Freeman; 2008.

24. E. D. Sverdlov. //Herald Russ. Acad. Sci 2006. V.76. №4, P.339.

25. Watson, J. D., Crick, F. H. C. A structure for deoxyribose nucleic acid. *Nature* 171, 737–738 (1953)

## APPENDIX A

In $E^3$ there are no closed (compact without edge) minimal surfaces, and the closure operation may be introduced by giving the metric (exterior in the given case), by bringing into correspondence with every pair of points x,y the distance ρ(x,y) between them. Catenoids, as well as helicoids locally isomorphic to them are closed subsets of $E^3$, which are the biggest minimal surfaces among all surfaces. The completeness of immersion of such surfaces in $E^3$ is provided by an exterior metric and by compactness of the subsets given on the surface. The latter implies that no single point not on a surface can be added to any subset. There are well developed methods of building complete minimal surfaces with zero index, embedded in $E^3$, using Weierstrass' representations, obtained using a 1-form, holomorphic on M and determined by the so called well meromorphic function. The latter include all meromorphic functions with finite number of zeros (or poles), all fractional rational, trigonometric and hyperbolic functions.

The Weierstrass representation itself is given by a pair fo functions (f, g), where f is holomorphic (as is $fg^2$), which does not become infinite for any values of the complex variable, is differentiable and defined everywhere is the complex planeкоторая не обращается в бесконечность ни при каких значениях комплексной переменной, and g is a meromorphic function (which does not only allow to define manifolds with isolated points, but also defines a mapping in $\mathbf{C}P^1 \cong S^3$). Zeros of the derivative function g correspond to points on a surface with zero Gaussian curvature (recall that the spherical torus on $S^3$ has zero curvature everywhere, dividing the giving sphere into two equal parts). If the function f is represented by a 1-form ω, and g is given on a Riemannian surface, then, under certain conditions, a global Weierstrass' representation is given which allows one to consider generalized minimal surfaces containing isolated points where regularity is broken. Note that the holomorphic 1-form within Weierstrass' representation may be extended to a meromorphic 1-form, which allows for putting such representations into correspondence with vector ones. This is possible because the surface may be defined both by 1-forms (corresponding to invariants of the system in question) and by vectors of the conjugated space.

It has been shown [6], that Weierstrass' representations allow one to define both a catenoid as well as a complete helicoid, and, in general, an associated family for some minimal surface M (for instance, of a helicoid or a catenoid) consists of locally isometric minimal surfaces (mutually incongruent, as a rule). Thus, the problem of constructing a surface with given properties turns out to be related both to constructing a coordinate cross for $S^3$ (in order to then use a cover over the bouquet $S^1 \cup S^2$), as well as with the construction of the plane torus ($T^2$ as a disk $D^2$).

Vectors characterizing discrete elements on a helicoidal surface can be put into correspondence with elements of the algebra $G_2$, which relate to automorphisms of lattices given below, put into correspondence with automorphisms of the lattice $E_8$. For instance, the second differential form of the surface M is given as a vector-valued bilinear form (with valued, in general, in $E^3$, and in the case being considered – in the appropriate algebra) in the following way. For a curve $\gamma(t)$ on M a (speed) vector is given with a value $\gamma(0)=P$ (so that the tangent plane $T_p$ is drawn thorough the point P), then the bilinear form $Q(v)$ (as a vector-valued form) can be given as a normal component of the acceleration vector normal to $T_p$. Thus, there appears a mechanism to construct minimal surfaces in correspondence with certain lattices, and, therefore, with algebras.

All non-planar non-congruent minimal surfaces form a one-parameter family of helicoids, or (for surfaces of revolutions) – a one-parameter family of catenoids. The parameters of such families may be chosen in the form of the pitch (h) for helicoids, and the radius (r) for catenoids. The difference for such configurations is that the generatrices are catenoids are the catenaries $\nu(u)$ in the plane XZ (so that with increasing u the Z-coordinate of points on $\nu(u)$ increases), while for the helicoid it is the Y axis (so that with increasing u the Y coordinate decreases). In a helicoid the median curvature along any generatrix equals zero, so that for any pair of generatrices there is a motion transforming one configuration into another and the helicoid into itself. Under certain conditions it is possible to build a configuration, unifying the helicoid and the catenoid, namely, a system of "summary" radius-vectors for both configurations for equal values of u.

## APPENDIX B

The fiber of the vector bundle is a real or complex vector space, and the structural groups G can be given by subgroups of linear, orthogonal or unitary subgroups. The structure of such fiber bundles f: E→M is determined by gluing functions over the intersections $U_{\alpha\beta}=U_\alpha \cap U_\beta$, and, consequently, by the mapping $T^{\alpha\beta}:U_{\alpha\beta}\to G$, so that $T^{\alpha\beta}=(T^{\alpha\beta})^{-1}$ and $T^{\alpha\beta}T^{\beta\nu}T^{\nu\alpha}=1$ in the intersection $U_{\alpha\beta\nu}=U_\alpha \cap U_\beta \cap U_\nu$ ($U_x$ – cover of the base M). Hence, over such fiber bundles, it is possible to perform all operations that preserve the above relationships, in particular, to use real or complex representations of the group G (as well as its other homomorphisms ρ:G→G) replacing the glue function $\rho(T^{\alpha\beta})= \rho BT^{\alpha\beta}$. The covers themselves are defined as fiber bundles with discrete fibers, in the sense that for the total space all complete preimages $F_y=f^{-1}(y)$ (y∈M) are discrete fibers. In a discrete manifold M (a discrete transformation group) for every point y∈ M, the orbit G(y) is a discrete set of points with a neightborhood U such that the images g(U) (g∈G) are either disjoint or coincide. The cover f: M→N is determined by free action of the discrete transformation group G, if for any point of the base y∈ N, the fiber $F_y=f^{-1}(y)$ is an orbit of the group (N=M/G is a regular or main fiber bundle with a discrete group G acting in M via diffeomorphisms of homomorphisms in general case). Given the connectedness of the manifolds N and M, the cover is called irreducible, and trivial if M≅N×F, where F is a discrete fiber. Considering vector bundles with linear action of a group G, the elements of a Lie algebra g may be viewed as a vector field, so that the section of the algebraic bundle E→M may be viewed as a vector field given on M. Hence in order to study such systems it is necessary to use polytopes and the fiber bundle formalism. The group action itself (given by linear transformations) on a vector spaces is a linear representation of such group.

The problem of constructing a given fiber bundle using automorphisms of the root lattice $E_8$ of the algebra $e_8$ (or the lattice isomorphic to it for the algebra of Cayley's octonions, or the module $2A_4$ for the Hurwitz group of quaternions), is reduced to building a cover (a discrete variant of fiber bundle) with certain properties. Where to conserve invariant properties of such transformation as well as to establish inheritance of locally-periodic properties (from prophase), one constructs a homogeneous space generated by the lattice $E_8$ and by a polytope on $S^3$, followed by the construction of a cell complex for $S^3$, as well as for the bouquet $S^1 \cup S^2$. In order to define locally-periodic properties when building cell complexes one uses mixes abelian groups containing torsionless subgroups (every element possesses finite periodicity) and torsion subgroups (using elements of unipotent subgroups not equal to unity). In the chain above, the connecting role is played by the $1^{st}$ and $2^{nd}$ differential forms with fixed properties, but given in different ways.

For such 1-connected covers we have $\pi_1(N,y_0) \cong \Gamma$, where $\Gamma$ is a transformation group M→M, so that the points of the inverse image $f^{-1}(y_0) = \{x_1, x_2...\}$ correspond to the monodromy representation of the cover $\sigma$. Thus, $\sigma$ is a homomorphism of the said fundamental group into the permutation group of points of the fiber that can be numbered by whole numbers. The subgroup $f_*\pi_1(M,x_j)$ of the group $\pi_1(N,y_0)$ consists of elements $\alpha \in \pi_1(N,y_0)$, for which the monodromy $\sigma(\alpha)$ leaves in place $x_j$, and is a normal subgroup of the group $\pi_1(N,y_0)$, and the group of monodromy itself is a factor-group $\pi_1(N,y_0)/f_*\pi_1(M,x_j)$ for any point $x_j \in F$. For the covering homotopy there is a unique dependence on M and N.

Let the points $y_i, y_j \in N$ have a neightborhood $U_j \in N$, so that every inverse image $f^{-1}(U_j)$ is represented by a union of non-intersecting domains. In covering the complete inverse image of a piecewise-smooth path $\gamma$ (with different ends $y_0$ and $y_1$) is diffeomorphic to the union of non-self-intersecting segments whose number equals the number of points in the fiber ($f^{-1}(y) \cong \gamma \times F$), so that every segment is projected in a diffeomorphic way under the mapping f onto a path $\gamma$ in the base. For covers of the form f: M→N with a discrete fiber $f^{-1}(y_0)$ for n≥2 there is an isomorphism $\pi_1(M,f_0) \cong \pi_1(N,y_0)$, so that there is a continuous mapping $D^3$→M with $\partial D^3 = F(S^2)$ and $s_0 \to f_0$ (a point, selected in F), sending $S^2$ into $f_0$ ($\pi_1(RP^2,f_0) \cong \pi_1(S^2,s_0)$ ).

Defining the base as a projective construction corresponding to a polytope, its edges may be viewed as the mentioned segments, and the Petrie polygon covering all its vertices as corresponding to the abovementioned non-self-intersecting path $\gamma$. In order for a homomorphism to take place (and a diffeomorphism in particular), it is necessary to define a fiber construction as a projective line, the polytope itself as a projective variety, determined by an appropriate incidence table, and the principal bundle space – as a cell manifold, put into correspondence with a polytope given on $S^3$. Recall that for a reductive group $G/B \cong P^1$, where B is a Borel subgroup.

## APPENDIX C.

If $K^n$ is a cell complex, then $\pi(K^n,S^n) = \pi(K^n,Z)$ determines homotopic classes of the map $K^n \to S^n$. An analogous situation takes place for fiber bundles, if the base is represented by a simplicial or cell complex. Using a 3D base, one can consider various sections (fields), allowing one to construct sections for the entire $M^3$. For instance, it is possible to consider handles of the type $H_1^3$ ($H_2^3$) = $D^1 \times D^2$ (as a manifold with edge of index 1), for which the boundary $dH_1^3$ has the form $dH_1^3 = (S^0 \times D^2) \cup (D^1 \times S^1)$. A corresponding manifold $K^3$ (with the edge $\partial K^3$) is constructed by gluing $K^3$ and the handle $H_1^3$ using the map f: $S^0 \times D^2 \to T_\varepsilon(S^0)$, where $T_\varepsilon(S^0)$ is a tube-like surface. Such diffeomeorphic constructions correspond to covers – discrete algebraic fiber bundles.

As shown in [], any smooth compact connected manifold $M^n$ is diffeomorphic to a union of handles $\{H_\lambda^n\}$, where $\lambda$ is the index $P_\lambda$ of a critical point (of the corresponding handle) of some Morse function on $M^n$. Correspondingly, to construct a surface one may use the following algorithm. It can be shown that on a smooth compact closed manifold $M^n$ there always is a Morse function with one maximum point (of index n) and one minimum point (of index 0), which is called regular if its critical values are partially ordered, namely, if equal or greater values correspond to points with equal or greater indices.

Any smooth compact closed manifold $M^n$ is diffeomorphic to a union of handles $\{H_\lambda^n\}$, where $P^\lambda$ are critical values of some Morse function with index $\lambda$, and every point $P^\lambda$ is in correspondence with some handle $H_\lambda^n$. Here $H_\lambda^n = D^\lambda \times D^{n-\lambda}$ and the boundary $dH_\lambda^n = (S^{\lambda-1} \times D^{n-\lambda}) \cup (D^\lambda \times S^{n-\lambda-1})$, so that for n=2 there are two variants to glue the handle $H_1^2$ to $H_0^2$, namely, $H_1^2 \cup H_0^2 \cong S^1 \times D^1$ (cylinder) и $H_1^2 \cup H_0^2$ (Mobius strip). If there exists a Morse function on $M^2$ with $x_0$ as the only minimum point, $x_1 \ldots x_N$ are points of index 1 ($H_0^2$ is homotopically equivalent to a 0-dimensional cell, $H_1^2$ is a handle) then starting with the point $x_2$ we glue a one-dimensional cell – we glue either $S^1 \times D^1$ or the Mobius strip, so that we traverse the point N we obtain a bouquet N of circles, each of which corresponds to a critical point. The last step consists in gluing the handle $H_2^2 \cong D^2$ (a cell can be identified with a fundamental polygon), which is realized according to identity transformation.

In order to consider all Morse functions possible on $M^n$, one should consider embedding of $M^n$ in $E^q$, because Morse functions, as a rule, can be identified with height functions ($h_l(x)$) possessing the following properties: a) the set $h_l(x)$ is in one-to-one correspondence with the points of the projective space $RP^q$ (or with a pairs of diametrically opposite points of the sphere $S^{q-1}$ in the case under consideration of embedding into $E^4$); б) the points $x_i \in M^n$ are critical for $h_l(x_0)$ only if the vector $l \perp Tx_0 M^n$ (in fact, a normal bundle is being considered for linear elements), so that, if Gaussian maps (embeddings) are defined as $M^n \to S^{q-1} \to E^q$ (in the case being considered $M^2 \to S^3 \to E^4$), then the critical points are not degenerate. There is also a simple way to construct the height functions: if one fixes (stable or zero) point $p \in E^4$, then it is possible to give on $M^2$ a smooth or piecewise-smooth function $L^2(x) = |p-x|^2$, where $x \in M^2$, and $|p-x|^2$ is squared length of the vector (p-x). In general, the set of such vectors does not coincide with the set of height functions, but for the case in question it is essential that for almost all points p such functions are not Morse functions. Restriction to a sphere and to root vectors simplifies the task, because the ring of invariants for the (Weyl) automorphisms group $E_8$ is such that the invariant of degree 2 is also the square of lengths of the vectors being used, which is invariant with respect to the mentioned group and is not contained in other invariants of the basis ring. Because for helicoids with the introduced exterior metric there is an embedding into $E^3$, it is possible to use already mentioned ways to introduce Morse functions on appropriate manifolds.

Embedding of the set of minimal geodesics, viewed as curves between points that correspond to identity transformations in the set of paths, is homeomorphic to the Grassmanian manifold $G^C_{2,1}$. Then the set itself is determined by the midpoints of such geodesics, coinciding with the intersection of the group SU(2) with the algebra su(2). For the group SU(2) one may consider embeddings of the circumference $S^1_0$ (1-parameter subgroup) and the 2D sphere $S^2_0$, whose equator is such a circumference $S^1_0$. A hemisphere of this sphere is identified with a 2D disk $D^2_0$, so that all constructions given here relate to minimal geodesic subsets. It is essential for further treatment that the group SU(2) is isometrically embedded into $S^7$ with the Killing's metric, invariant with respect to the right and left shifts for the group. The circumferences themselves $S^1_0 \subset SU(2) \subset S^7$ are the circumferences of the great circle of the sphere $S^7$, and the disks $D^2_0$ are the central plane sections of the sphere $S^7$ by a 3D plane going through the origin in $E^8$. Introducing $S^1_0 \subset T^1$ as part of the maximal torus in the group SU(2), invariants of the Weyl group of $E_8$ type, whose root lattice is considered, restricted to the sphere $S^7$, may be put into correspondence with $D^2_0$.

In building polytopes one, in fact, realizes the construction $S^3 \cong D^2_0 \times S^1$ as full tori ($D^2_0$ may be put into correspondence with $T^2$) and a homogeneous space (a symmetric space of the $1^{st}$ type) of the form SU(2)/U(1) using complex numbers U(1)={exp$i\varphi$}. Note the similarity between this situation with giving eigenvalues (of multiplicity 1) of the form $\exp 2\pi i m_j \varphi$ under Coxeter transformations (for Weyl groups). Any transformations of the disk while preserving local minimality can be reduced to rotating the disk around its boundary circumference $S^1_0$ while using automorphisms of the group SU(2). Further treatment is related to using gluing operations on $T^2$ do describe constructions of 3D manifolds.

The fiber bundle formalism allows one to tie the properties of simplicial (cell) complexes and their surfaces. In particular, faces are related to edges and their midpoints and vertices as singular points, because the solutions of the equation $\omega_E=0$ define in the main fibration (E) in the fiber (F) the families of "horizontal" n-dimensional directions. Here G-connection (G is the structural group of the fibration) defines a translation of fibers in the main fibration along any piecewise-smooth curve $\gamma(t)$ in the base, defined by shifts on G. At the same time, the analogous equation $\omega_F=0$ gives "horizontal" directions for points of the fiber in associated fibrations. If the fiber is a vector space and G acts linearly, then the elements of the Lie algebra g can be considered linear vector fields.

## APPENDIX D.

Any ordered set, for example, of elements of a field P, may be represented as a homogeneous polynomial in a variable x, with coefficients from P, by constructing the polynomial ring P[x], and vice versa – one may move from a polynomial ring to studying an ordered system of elements P, which is especially important for finite fields. When deriving non-integral axes one uses the fact that polynomials with non-integral coefficients, irreducible over the ring of integers, will also be irreducible over the field of rational numbers. Given a polynomial ring $P[x_1…x_n]$ of n variables over the field P (real field and 8 generators in the case of the lattice $E_8$), one may use the fact that this field is contained in some commutative ring L as a subfield. For instance in the ring of quaternions over the field of complex numbers, the addition of complex numbers may be interpreted as addition of vectors, so that the additive group of the field of complex numbers may be viewed as a 2D vector space over the field of real numbers. The minimal subring (L') in L consists of n elements $\alpha_1...\alpha_n$, through which all elements of L' and P may be expressed using addition (subtraction) and multiplication, hence we have that L' is isomorphic to the polynomial ring $P[x_1…x_n]$.

The rank of an abelian group (the number of infinite cyclic groups), as well as the orders of primary cyclic groups in any decomposition of an abelian group with a finite number of generators, are invariants of such group, so that its periodic part can be represented as a sum (by p), of cyclic elements, each of which is a direct sum. Further consideration of abelian groups is related to using solvable and one-dimensional nilpotent groups whose subgroups and homomorphic images are also nilpotent. The crux of the matter is that there is a one-to-one correspondence between torus groups and free discrete abelian groups, as well as between d-groups (in particular, tori) and torsionless finitely generated abelian groups. It is necessary to take into account that in contrast with commutative algebraic (or unipotent, which are not necessarily closed by themselves) groups, in reducible groups only those classes are closed that consist of semisimple elements (endomorphisms, when the minimal polynomial of the element x has n roots, namely, when it is diagonalizable over K).

An element x is nilpotent iff for some vector space over the field K there is an endomorphism $x^n=0$, where n is an integer, and 0 is the only eigenvalue of such endomorphism, so that 0 also simultaneously nilpotent. Finite p-groups are nilpotent and hence are representable by direct products of Sylow subgroups. A special place is occupied by the unipotent endomorphisms of the form $(1+ x_s^{-1}x_n)$, such that all its eigenvalues equal 1, where $x_n$ and $x_s$ are semisimple and nilpotent endomorphisms, respectively (the unity is the only endomorphism, which is both unipotent and semisimple). There is an essential difference in the behavior of semisimple and unipotent elements over the fields of characteristics 0 and p. Namely, if the behavior of semisimple elements in the said fields is identical, then unipotent elements, not equal to unity, for the field of characteristic zero have infinite order, and for char=p takes place $(x^p)^t$

The structure of the reductive group G is described by its standard parabolic subgroups which are in a one-to-one correspondence with the $2^l$ subsets of the base of roots $\Delta$ (l=rank G, in this case $2^l=4$), so that the subset I of the base $\Delta$ is in correspondence with the group $P_I=BW_IB$, where $W_I$ is generated by $\{\sigma| \alpha\in I\}$. Every parabolic subgroup is related to one and only one $P_I$ whose roots (with respect to T) are the elements $\Phi^+$, as well as the elements from $\Phi^-$, that are representable as Z-linear combination of elements of I.

If V is a closed unipotent subgroup of the group G, then $N=N_G(V)$ is a parabolic subgroup that determines $\{\alpha\in\Delta| U_\alpha\not\subset V,\}$. Every connected unipotent group lies in some Borel subgroup

(moreover, maximal unipotent subgroups of G are none other than unipotent radicals of Borel subgroups), every simple root of the base determines a maximum parabolic subgroup $P_i=P_{\Delta-\{\alpha i\}}$. Every element $\alpha_k$ ($k\neq i$) has a representation in $P_i$, which leaves fixed a vector $v_i \in V_i$ for a representation $G \to GL(V_i)$ with a line, whose stabilizer coincides with $P_i$. The group G itself acts on $v_i$ in a non-trivial way, so that we have a maximum vector of some (dominant) weight $\mu_i=\Sigma d_i\lambda_i$. Then if such dominant weight belongs to the group $X(T)$, there is an irreducible G-module with the higher weight mentioned above, as well as a unique $V(\lambda)$. If one uses a minimal vector, then any vector less than $\lambda$ is not dominant; at the same time is weight are conjugated with $\lambda$, and the dimension $V(\lambda)$ coincides with the number of elements of W-orbit for the weight $\lambda$. The zero vector is minimal, but it leads only to trivial 1D representations G. For root systems of the form $A_l$, the weights fundamental roots $\lambda_1...\lambda_l$, so that $\lambda_1$ is the highest weight of the natural (l+1)-dimensional representation.

## APPENDIX E.

Polytopes are given on $S^3$ as homogeneous spaces, hence we are using standard constructions of cell complexes for projective spaces (corresponding to the spheres $S^3$ and $S^2$), namely, $RP^3=D^3 \cup_f RP^2$, where $f:S^2 \to RP^2$ is a standard cover, or $RP^2=D^2 \cup_f RP^1$ where $f:S^1 \to RP^1$, and in the disks being used whose diametrically opposite points are glued over the boundary $S^2$ ($S^1$). Order to use the disks corresponding to the lattice $E_8$, cut of the sphere $S^7$ by a 3 hyperplane are used (for the space of paths when using $S^7$ as the space of the principal bundle for $SU(2)$).

The mapping f: $X \to Y$ is a fiber bundle, if it possesses the property of covering homotopy, namely, under the condition that any homotopy $K \times I \to Y$ (base) is covered by some homotopy $K \times I \to X$ (for all $0 \geq t \leq 1$), where a fixed point lying on some segment preserves the mentioned property; thus, there is one-to-one (multiplicative) dependence between a cover in X and motions of a point in the base Y.

In such constructions of principal interest are automorphisms groups (as groups of transformations) of manifolds in questions (algebraic ones, of the adjoint type). Chevalley has suggested a general construction for such groups, allowing to perform operations over arbitrary fields. It boils down to considering a root system (a basis) over the field **C** where all structural constants of the appropriate Lie algebra are whole numbers, and then automorphisms of the form $\exp ad x_\alpha$ (where $\alpha \in \Phi$ is a root system) are considered, which leave invariant the Z-envelope of the basis.

Correspondingly, such a group may be considered a matrix group over an arbitrary field полем (all automorphisms are inner). Because a non-zero vector field can be defined only on a torus (T), it becomes to recall the following properties of tori for various constructions. Using representations and one-parameter subgroups of the form $\lambda$: multiplicative group $G_m \to T$ means that such groups in projective constructions leave fixed the points $\lambda(0)$ and $\lambda(\infty)$. Here the mapping $0 \to \infty$, and, consequently, $\lambda(0)[v] = \lambda(\infty)[v] \to T \subset GL(V)$, leaves fixed the points $[v]$. But in our consideration it is essential that singular and cell simplicial homologies coincide with cell homologies, hence in considering equivalent spaces it suffices to use simplicial cell spaces with a given homology (когомологией).

Chevalley constructions generate not just the sets of matrices, but also corresponding division algebras. There is a natural generalization of such matrix algebras, namely, if the roles of new subsets and their corresponding subalgebras are taken by linear combinations from a given set of matrices (from the given set) of the same order with coefficients from the field *κ*. Furthermore, any automorphism of the direct sum of irreducible matrix algebras is generated by some non-singular matrix, and if elements of the center are not involved, such automorphism is also inner. Because, upon the action of the group G as an automorphism group, its algebra g acts as a differential algebra, use of diffeomorphisms and of algebraic representations considered above becomes a determining factor in the construction of covers within a local approach, in particular, when one considers local phase transformations/

In order to define an exterior form ω on M in $E^{2n}$, an orthonormal basis $e_1…e_n$ can be chosen where $ω=λ_1ω_1+…+λ_{2n-1}ω_{2n-1}\cap ω_{2n}$, and $λ_1..λ_{2n}$ are non-negative integers, $ω_1...ω_{2n}$ is the basis conjugate to the one given above. In order to define vector manifolds, polytopes in $E^4$ are constructed (as homogeneous spaces) on $S^3$, generated (by algebraic fibrations) by the root lattice $E_8$ of the algebra $e_8$. Then the so-called coordinate cross is singled out on $S^3$, allowing one for $S^3 = S^2 \cup S^1$ to consider covers as discrete fibrations for different constructions, including Chevalley's constructions.

Twisting the root system Φ of the type $D_4$ using the $3^{rd}$ order symmetry elements of the Coxeter graph gives a system associated with the root system of the type $G_2$, which allows one to put them in correspondence their root subgroups. Any large abelian subgroup of the unipotent subgroup $U$ of Chevalley's group $G(K)$ of classical type over a finite field is conjugated $G(K)$ with a normal large abelian subgroup from $U$. A commutative subset of the root system Φ is a subset $Ψ \in Φ$, for any of whose roots we have $r, s \in Ψ$, and their sum $r+s$ does not lie in Φ. For the Chevalley construction it is possible to put into one-to-one correspondences specific subsets of roots (taking into account relatedness of root systems of types $D_4$ and $G_2$) with abelian groups used for cell constructions.

For minimal surfaces their stability is determined with respect to change in their areas (as well as the volume within it) under small deformations. Stability is characterized by the index Ind M, which corresponds to the number of ways that the area of surface can be changed. Thus, if it is not zero, the surface M is unstable. Further consideration should take into account that the indices of all periodic minimal surfaces (In particular, Riemann-Shwartz's, complete Sherk's surface, infinite helicoid) are infinite and unstable in the sense above.

However, there are well-developed methods of building complete minimal surfaces of zero index, fully immersed in $E^3$, using the Weierstrass representation obtained using 1-form, holomorphic on M and defined by the so-called good meromorphic function. The latter include all meromorphic functions with finite number of zeros (or poles), all fractional-rational, trigonometric, and hyperbolic functions.

## APPENDIX F.

In addition to lattices, given over integral rings, lattice are possible over rings of algebraic numbers (I), containing irrational elements. These lattices are everywhere dense in $K^n$ (where K=R,**C**,H) and, therefore, do not form discrete lattices (they are used in consideration of irrational cuts in derivation of quasicrystals), but may be used to construct lattice packings by mean of renormalization of vector norms and the quadratic forms related to them. In those cases when I is not a main ideal domain, it is possible to define a lattice over Z, which will not be generated by n elements. The renormalization is performed by replacing the bilinear form for a Hermitian form and using a homomorphism whereby all elements of the ring map into elements of the field. For example, it is possible to construct a 8-dimensional lattice Z[ξ], where $ξ=e^{πi/4}=(1+i/\sqrt{2})$ by using elements of the field from $F_9 \times F_9$ (thus, in fact, the incidence table 9×9 is used).

For such construction, the roots of $8^{th}$ degree of 1 (with addition of a selected point) are considered as elements of the field. Then the elements from Z[ξ] with norm [2] are mapped into an element with norm [-1] from $F_9 \times F_9$. By analogy, elements with norm [4] are mapped into elements with norm -1, and elements with norm 6 (three times) – into nonzero elements with norm [0]. One uses the mapping [3] with the norm of the form N: $F_9 \times F_9 \to F_3$ with subsequent setting up of a correspondence between these elements and the 81 class of conjugated elements of the set $D_4/3D_4$, which finally leads to the root polytope {96}. In such case the polyhedron whose vertices coincide with 24 minimal vectors of the lattice $D_4$ may be represented by elements from Z[ξ].

A transition to using Hurwitz's whole numbers is related to the fact that the group of quaternions $|q|^2=1$ (restriction to a unit sphere) describes rotations in $E^3$ (and is as well in correspondence with reflection groups), as well as with a necessity to preserve the lattice property, and, therefore, a one-to-one link with the pre-phase $E_8$. Recall that $Q=\{x^2, x \in F_q\}$, $Ω=PL(q)$, $N'=N/\{∞\}, N=Ω/Q$. Thus, for finite projective constructions it is possible to use the mappings $E^3 \cup \{∞\} \to S^2$ provided that nonzero vector fields are given and with transition from the manifold N to the manifold N′. It is the 24-element construction that is used both in building the polyhedron {3,4,3} and in the subsequent setting up of a correspondence with the Hurwitz group of unit quaternions, which in turn is in correspondence with the 24 –element scale invariant subgroup of the group SU(2), and is also used when considering local (homotopically equivalent) phase transitions with fixed number of vertices. When considering the set $A_2/3A_2$ for $G_2$ we have nine classes (including zero).

A lattice of such representations (as well as the group exp p, related to unipotent representations), corresponding to $E_8$, may be constructed using, for example, the ring of Hurwitz's integers (*H*-lattice) – 24 quaternions of norm 1. It can be considered in its turn as a multiplicative group $2A_4$ ($A_4 \oplus A_4$), where $A_4$ is the alternating group of degree 4. In the case of *H*-lattice acting in $\mathbf{H}^1$ it is also possible to construct (using the $2A_4$ module) a L-lattice, whose real lattice is going to be $D_4$. Then every non-trivial element of $2A_4$ acts without fixed points (there are 22=2·11 of them in the mentioned ring or, taking into account the two identity transformations, 20=2·10). Formally the doubling operation may be viewed as the correspondence of every critical point on $M^2$ to two critical points on a tube surface.

The Mathieu group $M_{24}$ is related to the invariants of $E_8$, in particular with the invariant 24 (exponential 23). This determines the existence of such its subgroups as $L_2(23)$, $L_2(9)$, $L_2(7)$, and $M_{11}$. The group $M'_{10}$= $PSL_2(9)$ sends the 10 points $v_i$ into 10 points of a projective plane, on which the elements of $M'_{10}$ act as fractional linear transformations with orbits like [1,1,1,1,20] and [6,6,6,6]. The Steiner systems S(5,8,24), S(4,7,23), S(5,6,12), S(4,5,11) are related to subgroups of $M_{24}$ as well as the alternating group $2A_4$ ($A_4 \oplus A_4$), which may be viewed as a multiplicative analog of the ring of Hurwitz's whole numbers (unitary quaternions), allowing to define the lattices $D_4$ and $E_8$. In order to define $M_{24}$ one uses $L_2(23)$ adding the transformation $x \rightarrow 9x^3$ ($x \in N$), hence the stabilizer of three points is transitive on the remaining 21 (analogous situation takes place with 4 or 5 points), and the group $M_{24-\kappa}$ may be defined as the stabilizer of any κ-element subset of Ω for κ≤5.

A.T. Fomenko, A.A. Tuzhilin. Elements of the geometry and topology of minimal surfaces in three-dimensional space. Translations of Mathematical Monographs, 1992. v.93.

B.L. Dubrovin, S.P. Novikov, A.T. Fomenko. Modern Geometry Éditorial URSS, Moscow, 2001. [in Russian].

J.H. Conway, N.J.A Sloane. Sphere Packing, Lattice and Groups. New York, Berlin, Heidelberg, London, Paris, Tokyo: Springer-Verlag 1988.

J. Humphreys. Linear Algebraic Groups. New York, Berlin: Springer-Verlag. 1975.

A.G. Kurosh. Theory Groups. Moscow. 1953. [in Russian].